\documentclass[aps,prl,twocolumn,superscriptaddress,showpacs]{revtex4}

\usepackage{graphicx}

\begin{document}


\title{Resonant Enhancement of Inelastic Light Scattering in the Fractional Quantum Hall Regime at $\nu=1/3$}


\author{C.F. Hirjibehedin}
\affiliation{Department of Physics, Columbia University, New York, NY
10027} \affiliation{Bell Labs, Lucent Technologies, Murray Hill, NJ
07974}

\author{Irene Dujovne}
\affiliation{Department of Appl. Physics and Appl. Mathematics,
Columbia University, New York, NY  10027} \affiliation{Bell Labs,
Lucent Technologies, Murray Hill, NJ 07974}

\author{I. Bar-Joseph}
\affiliation{Department of Condensed Matter Physics, The Weizmann
Institute of Science, Rehovot 76100, Israel}

\author{A. Pinczuk}
\affiliation{Department of Physics, Columbia University, New York, NY
10027} \affiliation{Bell Labs, Lucent Technologies, Murray Hill, NJ
07974} \affiliation{Department of Appl. Physics and Appl.
Mathematics, Columbia University, New York, NY  10027}

\author{B.S. Dennis}
\affiliation{Bell Labs, Lucent Technologies, Murray Hill, NJ 07974}

\author{L.N. Pfeiffer}
\affiliation{Bell Labs, Lucent Technologies, Murray Hill, NJ 07974}

\author{K.W. West}
\affiliation{Bell Labs, Lucent Technologies, Murray Hill, NJ 07974}

\date{June 25, 2003}

\begin{abstract}
Strong resonant enhancements of inelastic light scattering from
the long wavelength inter-Landau level magnetoplasmon and the
intra-Landau level spin wave excitations are seen for the
fractional quantum Hall state at $\nu = 1/3$. The energies of the
sharp peaks (FWHM $\lesssim 0.2meV$) in the profiles of resonant
enhancement of inelastic light scattering intensities coincide
with the energies of photoluminescence bands assigned to
negatively charged exciton recombination. To interpret the
observed enhancement profiles, we propose three-step light
scattering mechanisms in which the intermediate resonant
transitions are to states with charged excitonic excitations.
\end{abstract}



\maketitle


Inelastic light scattering is a key method for the study of
collective excitation modes in quantum Hall (QH) liquids
\cite{Pinczuk1988,Pinczuk1992,Pinczuk1993,Davies1997,Schuller1997,Pellegrini1998,Eriksson1999,Kang2001,Kulik2001,Dujovne2003b}.
Resonant enhancement of the scattering cross-section is here crucial
because the non-resonant scattering intensity in 2D electron systems
is too small to be observable
\cite{Burstein1979,Burstein1980,Pinczuk1988,Pinczuk1994,Platzman1994,DasSarma1999}.
The enhancement of the scattering cross-section also plays a role in
degenerate four-wave mixing \cite{Fromer2002} and coherent optical
oscillation \cite{Bao2002} studies. In the presence of weak residual
disorder, breakdown of wavevector conservation in light scattering
processes offers direct access to critical points in the dispersions
of charge and spin excitations of QH liquids
\cite{Pinczuk1988,Pinczuk1992,Marmorkos&DasSarma1992,Pinczuk1993,Davies1997,Eriksson1999,Kang2001}.
However, only brief attention has been given to the resonance
profiles and the details of the underlying enhancement mechanisms of
the observed collective excitations.
\par

Here we present resonance enhancement profiles in the fractional
QH (FQH) regime and discuss their implications, particularly in
relation to optical transitions that have been the focus of much
recent work
\cite{MacDonald1992,Wang1992,Apalkov1993,Kheng1993,Wojs1995,Finkelstein1995,Shields1995,Palacios1996,Gekhtman1997,Wojs1999,Wojs2000,Yusa2001,Schuller2001,Szlufarska2001,Nickel2002,Broocks2002}.
We focus our attention on the resonance enhancement profiles for
inelastic light scattering when the electron system is in the
major FQH state with $\nu=1/3$, where $\nu=n h c / e B$ is the
Landau level filling factor for areal density $n$ and
perpendicular magnetic field $B$. We consider the enhancement
profiles of the inter-Landau level magnetoplasmon (MP) mode at the
cyclotron energy $\omega_c$ and the intra-Landau level spin wave
(SW) at the Zeeman energy $E_{z}$. In the MP mode only the
electron Landau level index changes and in the SW mode there is
only reversal in spin orientation of electrons and/or
quasiparticles of quantum liquids. Both collective modes occur at
$q\rightarrow 0$, so that the impact of breakdown of wavevector
conservation is ignored in a lowest order description.
\par

We find strong enhancement of the light scattering intensity within a
small range of incoming and outgoing photon energies. These
enhancements coincide with optical transitions assigned to
recombination of negatively charged excitons ($X^-$)
\cite{Kheng1993,Shields1995,Wojs2000,Yusa2001}. The large resonance
enhancements reported here enable light scattering studies at the
very low light power densities required to achieve temperatures in
the milliKelvin regime under experimental conditions
\cite{Kang2000,Dujovne2003a,Hirjibehedin2003}. In a qualitative
interpretation, we propose three-step light scattering mechanisms
that interpret the resonance enhancements with intermediate
transitions to states with $X^-$ excitations
\cite{Burstein1979,Burstein1980,Pinczuk1988,Pinczuk1994,Platzman1994,DasSarma1999}.
\par

The links between resonant enhancements of light scattering and the
$X^-$ excitations provides new venues for the study of the properties
of $X^-$. Studies of resonant enhancement of light scattering by
modes seen due to breakdown of wavevector conservation in the FQH
regime \cite{Kang2001,Dujovne2003a} will be the subject of future
work. Well-defined polarization selection rules, which are seen in
some resonances, also offer new venues for exploring the properties
of $X^-$ transitions.
\par

\begin{figure}
\includegraphics{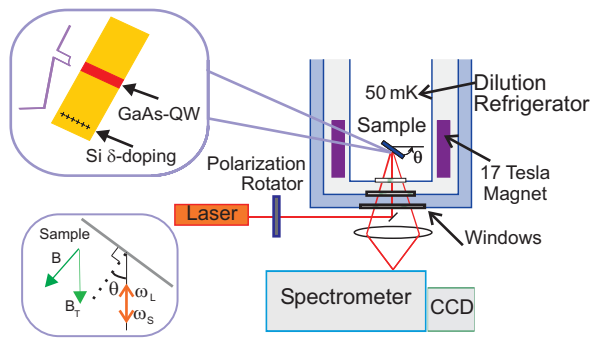}
\caption{\label{fig:figure1} A schematic of the experimental setup
and measurement configuration. Linearly polarized light from a
tunable laser is passed through a polarization rotator and directed
onto the sample through windows. The scattered light is focused into
a spectrometer and the dispersed light is detected by a CCD. The
upper inset shows the sample design and a sketch of the conduction
band. The lower inset shows the backscattering geometry.}
\end{figure}

The 2D electron systems studied here are formed in $250\AA$- and
$330\AA$-wide asymmetrically doped GaAs single quantum wells,
referred to as samples A and B. The electron densities for the two
samples are $8.5 \times 10^{10}cm^{-2}$ and $5.4 \times
10^{10}cm^{-2}$ and the low temperature mobilities are $3.0 \times
10^6 cm^2/Vs$ and $7.2 \times 10^6 cm^2/Vs$, respectively. As shown
in the schematic in Fig. \ref{fig:figure1}, the samples are mounted
on the cold finger of a dilution refrigerator with a base temperature
of $50mK$. This is inserted into the cold bore of a $17T$
superconducting magnet. The samples are mounted with the normal to
the surface at an angle $\theta$ from the magnetic field, making the
component of the field perpendicular to the 2D electron layer $B =
B_T cos \theta$ for a total applied field $B_T$.
\par

As shown in the schematic in Fig. \ref{fig:figure1}, light scattering
measurements are performed through windows for direct optical access.
The energy of the linearly polarized incident photons is tuned close
to the fundamental optical gap of the GaAs well and the power density
is kept below $10^{-4} W/cm^2$. The samples are measured in a
backscattering geometry, making an angle $\theta$ between the
incident/scattered photons and the normal to the sample surface. The
wavevector transferred from the photons to the 2D system is $k=(2
\omega_L / c) sin \theta$. For typical measurement geometries,
$\theta < 60^{\circ}$ so that $k \leq 1.5 \times 10^5 cm^{-1} \ll
1/l_0$, where $l_0 = (\hbar c / e B)^{1/2}$ is the magnetic length.
Scattered light is dispersed by a Spex 1404 double spectrometer with
holographic master gratings that reduce the stray light and provide a
resolution as small as $20 \mu eV$. The response of the spectrometer
is linearly polarized, so that spectra can be taken with the linear
polarization of the incident photons parallel (polarized) or
perpendicular (depolarized) to the detected scattered photons'
polarization.
\par

\begin{figure}
\includegraphics{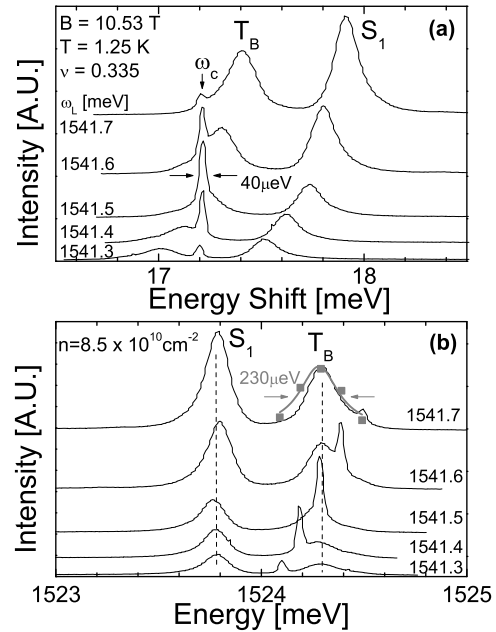}
\caption{\label{fig:figure2} Spectra at various $\omega_L$ for $\nu =
1/3$ in sample A (after Ref \cite{PinczukSST1994}), translated
vertically for clarity. (a) Spectra are plotted as intensity vs.
energy shift below $\omega_L$. The mode at a constant shift of
$17.2meV$ labelled $\omega_c$ is light scattering from the MP
excitation. (b) The same spectra are plotted as intensity vs.
absolute photon energy. The peaks labelled $S_1$ and $T_B$ are
identified as luminescence from $X^-$ excitations, as discussed in
the text. Grey dots overlayed on the $1541.7meV$ spectrum are the
intensity of the light scattering from the MP mode vs. $\omega_S$,
and the grey line is a fit using Eq. \ref{eq:MP_intensity}.}
\end{figure}

Figure \ref{fig:figure2}a shows spectra at $\nu=1/3$. In these
spectra, the collective excitation energy is measured as a shift
below the laser energy $\omega_L$. The sharp peak labelled
$\omega_c$, found at a constant shift of $17.2meV$ for various
$\omega_L$, is inelastic light scattering from the long wavelength
(i.e. at wavevector $q\rightarrow 0$) MP. This mode is fixed by
Kohn's theorem at the cyclotron energy $\omega_c = e B / m^* c$. Two
additional features are also seen at shifts from the laser that
depend on laser photon energy. These are more easily understood in
Fig. \ref{fig:figure2}b, where the spectra are plotted on an absolute
energy scale. The features at $1523.7meV$ and $1524.3meV$ are
luminescence from optical transitions identified as the singlet
($S_1$) and bright triplet ($T_B$) $X^-$ recombination
\cite{Kheng1993,Shields1995,Wojs2000,Yusa2001}. From Fig.
\ref{fig:figure2}, it is clear that for the excitation at $\omega_c$
the light scattering process is resonant when the outgoing photon
energy $\omega_S$ coincides with the $T_B$ transition.
\par

Overlap of $\omega_L$ or $\omega_S$ with an $X^-$ transition can
cause striking increases in the light scattering intensities by
excitation modes of liquids in the FQH regime. We see from Fig.
\ref{fig:figure2}a that the intensity of the MP mode in sample A
becomes significantly larger when $\omega_S$ overlaps $T_B$.
Remarkably, the intensity of the scattered light at the peak of this
outgoing resonance can be larger than that of the luminescence. In
Fig. \ref{fig:figure2}b we show the resonance enhancement profile for
the MP mode as a function of $\omega_S$ overlaid on the luminescence.
The lineshape of the outgoing resonance profile of light scattering
intensity corresponds very closely to that of the $T_B$ luminescence,
and both are peaked at $1524.28meV$. The $T_B$ luminescence lineshape
has a FWHM of $0.17meV$, which is narrower than the FWHM of $0.23meV$
for the resonance profile.
\par

\begin{figure}
\includegraphics{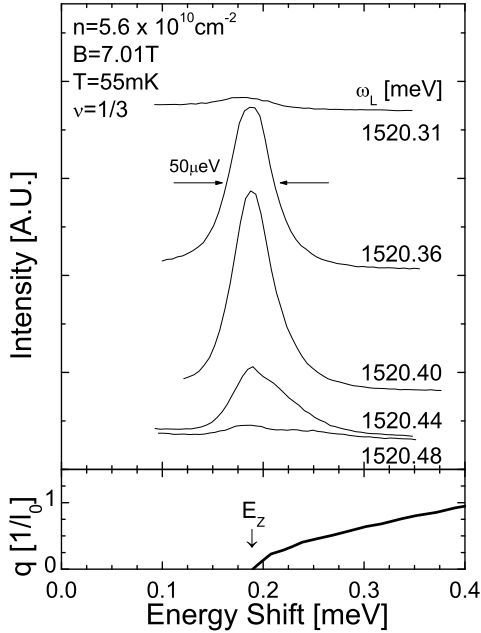}
\caption{\label{fig:figure3} (a) Spectra at various $\omega_L$ for
$\nu = 1/3$ in sample B, translated vertically for clarity. Spectra
are plotted as intensity vs. energy shift below $\omega_L$. (b)
Calculation of SW dispersion \cite{Nakajima&Aoki1994} including a
scaling by a factor of $0.59$ to incorporate finite width corrections
\cite{Kang2001,Hirjibehedin2003}.}
\end{figure}

Figure \ref{fig:figure3}a shows light scattering spectra from the
long wavelength intra-Landau level spin wave (SW) excitation in
sample B. The energy of this mode is fixed by Larmor's Theorem at the
Zeeman energy $E_z = g \mu_B B_T$, where $g$ is the Lande g-factor in
GaAs and $\mu_B$ is the Bohr magneton \cite{Pellegrini1998}. The SW
is resonant throughout a broad range of $\omega_L$ and $\omega_S$. In
the results reported here the focus of our current analysis is on the
strongest and lowest energy resonance, where the peak intensity is at
least a factor of two larger than in other resonances. One striking
difference between these spectra and those for the $\omega_c$
resonance is that no luminescence is seen when $\omega_L$ overlaps or
is below the energy of the neutral exciton ($X$) optical
recombination energy. It is possible that photoexcited states are
unable to relax and then emit if the exciting photon energy is below
a certain threshold. This is consistent with our observations that
the luminescence appears for higher energy $\omega_L$.
\par

As seen in Fig. \ref{fig:figure3}a, the SW has a relatively narrow
$FWHM = 50 \mu eV$. However, its lineshape is asymmetric because of
an enhanced tail on the high energy side. In Fig. \ref{fig:figure3}b
we show the SW dispersion $\omega_{SW} = E_z + \alpha
\Delta_{SW}(q)$, where $\Delta_{SW}(q)$ is calculated exactly for
finite sized systems \cite{Nakajima&Aoki1994} and $\alpha = 0.59$ is
a constant scaling factor to account for finite width effects
\cite{Kang2001,Hirjibehedin2003}. Breakdown of wavevector
conservation from residual disorder can activate larger wavevector SW
excitations. As seen in Fig. \ref{fig:figure3}b these occur above
$E_z$, causing the observed asymmetry \cite{Pinczuk1998}.
\par

\begin{figure}
\includegraphics{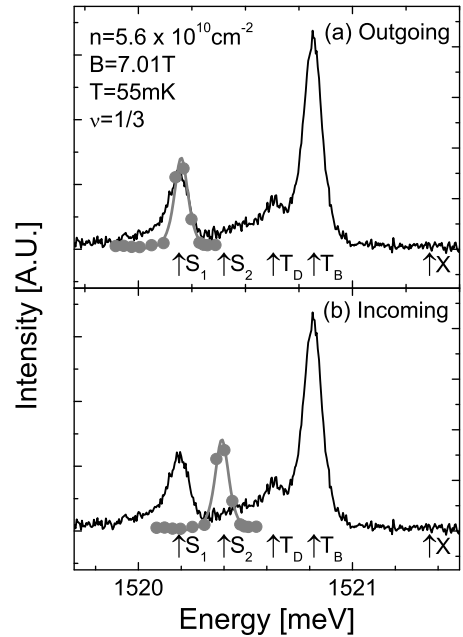}
\caption{\label{fig:figure4} Luminescence intensity vs. absolute
energy for sample B taken with $\omega_L = 1525.02meV$. Arrows at the
bottom of each panel label the transitions to which the features are
assigned. The overlayed grey dots are the resonance enhancement
profile of the SW vs. (a) $\omega_S$ and (b) $\omega_L$. The
intensities have been scaled by a constant to match the intensity of
the $S_1$ luminescence.}
\end{figure}

In Fig. \ref{fig:figure4}, we compare the SW resonance enhancement
profile to the luminescence lineshape. Assignment of the two singlet
($S_1$ and $S_2$), dark triplet ($T_D$), and bright triplet ($T_B$)
luminescence features is made based on the arguments presented in
Refs. \cite{Kheng1993,Shields1995,Wojs2000,Yusa2001}. The $S_2$
transition has a particularly low intensity, but is more clearly seen
in other spectra (not shown). The peak in the resonance profile has a
$FWHM <0.1meV$, which is slightly sharper than the luminescence
linewidths. From Fig. \ref{fig:figure4}a, we find the peak in
scattering intensity occurs when $\omega_S$ coincides with the
feature assigned to the $S_1$ luminescence transition. The incoming
resonance shown in Fig. \ref{fig:figure4}b indicates that the peak
also overlaps the feature assigned to the $S_2$ luminescence.
\par

\begin{figure}
\includegraphics{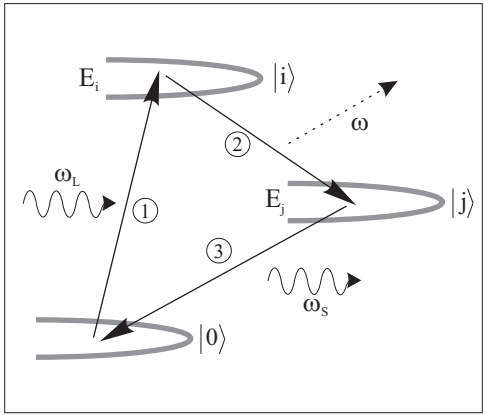}
\caption{\label{fig:figure5} Diagrammatic representation of
third-order resonant light scattering process described by Eq.
\ref{eq:intensity}. The numbers indicate the time ordering of the
processes.}
\end{figure}

The simplest mechanism for wavevector conserving light scattering
processes that exhibit both incoming and outgoing resonances is a
third-order process. The intensity of this process, which is
described in the diagram shown in Fig. \ref{fig:figure5}, can be
written within time-dependent perturbation theory as
\cite{Pinczuk1988,Danan1989}
\begin{equation}
I(\omega) \sim \left| \sum_{i,j}\frac{\langle 0 | H_{\gamma}| j
\rangle \langle j | H_{int} | i \rangle \langle i | H_{\gamma} | 0
\rangle}{(\omega_S - E_{j})(\omega_L - E_{i})} \right|^2,
\label{eq:intensity}
\end{equation}
where the intermediate states $| i \rangle$ and $| j \rangle$ have
energies $E_i$ and $E_j$ above the ground state respectively. In the
first step, the incoming photon $\omega_L$ is annihilated and the
system goes from state $|0 \rangle$ to an intermediate state $|i
\rangle$ via a virtual transition through the electron-photon
interaction $H_{\gamma}$. The system then goes from $|i \rangle$ to
$|j \rangle$ through the $H_{int}$ interaction between the electrons
and the collective excitations of the 2D system. In this second step
there is emission of a collective excitation at $\omega$. In the
third and final step, the electron system returns to state $| 0
\rangle$ with the additional collective excitation at $\omega$ and
emits a photon at energy $\omega_S = \omega_L - \omega$. The
quasiparticle properties of the electron liquids are here
incorporated into $H_{int}$.  In the specific case of the SW we note
that the excitation is a collective mode associated with
spin-reversed transition of composite fermions (CFs)
\cite{Nakajima&Aoki1994,Mandal2001}. Here $H_{int}$ must arise from
the residual interactions between composite fermions.
\par

The peaks in the resonance profiles shown in Figs. \ref{fig:figure2}
and \ref{fig:figure4} coincide with luminescence transitions of
$X^-$. From this, we infer that the intermediate states of the
third-order light scattering mechanism are excitations of the ground
state that can include $X^-$ transitions and a collective excitation.
Enhancement of light scattering in the FQH regime is then directly
linked to $X^-$ transitions, and their symmetry properties will
determine the selection rules.
\par

Light scattering from the MP mode is seen in Fig. \ref{fig:figure2}
to have an outgoing resonance with the $T_B$ state. Assuming there is
no double resonance condition (simultaneous incoming and outgoing
resonances), the scattering intensity in this range can be written as
\begin{equation}
I(\omega) \sim \frac{1}{(\omega_S - E_{T_B})^2 + \Gamma^2},
\label{eq:MP_intensity}
\end{equation}
where $\Gamma$ is a broadening of the transition. The best fit of Eq.
\ref{eq:MP_intensity} to the resonance profile of the MP mode with
$\Gamma = 0.11meV$ is shown in Fig. \ref{fig:figure2}. The FWHM is $2
\Gamma = 0.22meV$, which is consistent with the width of the $T_B$
transition as measured by the luminescence linewidth.
\par

From Fig. \ref{fig:figure4}, the SW mode shows an outgoing resonance
but no incoming resonance with the $S_1$ state. Assuming the product
of matrix elements in Eq. \ref{eq:intensity} is the same under
interchange of states $i$ an $j$, we would expect an incoming and
outgoing resonance at close to the same intensity. Because the SW
energy at $1/3$ in sample B is close to the spacing of the $X^-$
transitions shown in the luminescence of Fig. \ref{fig:figure4}, it
is possible to have a double resonance condition, in which the
incoming and outgoing resonances for two states coincide. For the SW
resonance shown in Fig. \ref{fig:figure4}, the spacing of the $S_1$
and $S_2$ transitions is close to $E_z = 0.19meV$. The resonance
lineshape from Eq. \ref{eq:intensity} can then be written as
\begin{equation}
I(\omega) \sim \frac{1}{((\omega_S - E_{S_1})^2 + \Gamma^2)((\omega_L
- E_{S_2})^2 + \Gamma^2)}, \label{eq:SW_intensity}
\end{equation}
where both levels are assumed to have the same broadening. The best
fit of Eq. \ref{eq:SW_intensity} to the resonance profile of the SW
mode with $\Gamma = 0.06meV$ is shown in Fig. \ref{fig:figure4}. The
FWHM of the lineshape is slightly wider than this at $0.08meV$. This
is sharper than the linewidth of $S_1$ or $S_2$, indicating that the
luminescence transition is broadened by additional mechanisms that do
not contribute to the light scattering process.
\par

Large resonance enhancements of the light scattering intensities such
as those reported here are crucial in the study of FQH liquids.
Significant examples are found in recent work that examines the
important low-lying excitations that manifest key characteristics of
the liquids, as in recent studies between major fractions in the
sequence $\nu = p / (2 p + 1)$ \cite{Dujovne2003a,Dujovne2003b} and
FQH states in the sequence $\nu = p / (4 p \pm 1)$
\cite{Hirjibehedin2003} for integer $p$. The excitations measured in
these studies typically have low energy ($<0.5meV$) and require
temperatures reaching below $100mK$. Because of the small ($<1meV$)
spacing between various $X^-$ transitions, double resonances similar
to those seen for the SW are possible for many low-energy excitations
in the FQH regime. With such resonances, studies can be performed
using the very low light power densities required to obtain very low
temperatures.
\par

In summary, we find strong resonance enhancement of inelastic light
scattering in the FQH state at $\nu = 1/3$.  Enhancement profiles of
long wavelength MP and SW modes show incoming and outgoing resonances
that coincide with $X^-$ optical transitions seen in luminescence.
The enhancements are interpreted by three-step light scattering
mechanisms in which the intermediate states are excitations of the
ground state with $X^-$ excitations.
\par



\begin{acknowledgments}
This work was supported in part by the Nanoscale Science and
Engineering Initiative of the National Science Foundation under
NSF Award Number CHE-0117752, by a research grant of the W. M.
Keck Foundation, and by the Israeli Academy of Science.
\end{acknowledgments}



\begin{thebibliography}{42}
\expandafter\ifx\csname
natexlab\endcsname\relax\def\natexlab#1{#1}\fi
\expandafter\ifx\csname bibnamefont\endcsname\relax
  \def\bibnamefont#1{#1}\fi
\expandafter\ifx\csname bibfnamefont\endcsname\relax
  \def\bibfnamefont#1{#1}\fi
\expandafter\ifx\csname citenamefont\endcsname\relax
  \def\citenamefont#1{#1}\fi
\expandafter\ifx\csname url\endcsname\relax
  \def\url#1{\texttt{#1}}\fi
\expandafter\ifx\csname
urlprefix\endcsname\relax\def\urlprefix{URL }\fi
\providecommand{\bibinfo}[2]{#2}
\providecommand{\eprint}[2][]{\url{#2}}

\bibitem[{\citenamefont{Pinczuk et~al.}(1988)\citenamefont{Pinczuk, Valladares,
  Heiman, Gossard, English, Tu, Pfeiffer, and West}}]{Pinczuk1988}
\bibinfo{author}{\bibfnamefont{A.}~\bibnamefont{Pinczuk}},
  \bibinfo{author}{\bibfnamefont{J.~P.} \bibnamefont{Valladares}},
  \bibinfo{author}{\bibfnamefont{D.}~\bibnamefont{Heiman}},
  \bibinfo{author}{\bibfnamefont{A.~C.} \bibnamefont{Gossard}},
  \bibinfo{author}{\bibfnamefont{J.~H.} \bibnamefont{English}},
  \bibinfo{author}{\bibfnamefont{C.~W.} \bibnamefont{Tu}},
  \bibinfo{author}{\bibfnamefont{L.}~\bibnamefont{Pfeiffer}}, \bibnamefont{and}
  \bibinfo{author}{\bibfnamefont{K.}~\bibnamefont{West}},
  \bibinfo{journal}{Phys. Rev. Lett.} \textbf{\bibinfo{volume}{61}},
  \bibinfo{pages}{2701} (\bibinfo{year}{1988}).

\bibitem[{\citenamefont{Pinczuk et~al.}(1992)\citenamefont{Pinczuk, Dennis,
  Heiman, Kallin, Brey, Tejedor, Schmitt-Rink, Pfeiffer, and
  West}}]{Pinczuk1992}
\bibinfo{author}{\bibfnamefont{A.}~\bibnamefont{Pinczuk}},
  \bibinfo{author}{\bibfnamefont{B.~S.} \bibnamefont{Dennis}},
  \bibinfo{author}{\bibfnamefont{D.}~\bibnamefont{Heiman}},
  \bibinfo{author}{\bibfnamefont{C.}~\bibnamefont{Kallin}},
  \bibinfo{author}{\bibfnamefont{L.}~\bibnamefont{Brey}},
  \bibinfo{author}{\bibfnamefont{C.}~\bibnamefont{Tejedor}},
  \bibinfo{author}{\bibfnamefont{S.}~\bibnamefont{Schmitt-Rink}},
  \bibinfo{author}{\bibfnamefont{L.~N.} \bibnamefont{Pfeiffer}},
  \bibnamefont{and} \bibinfo{author}{\bibfnamefont{K.~W.} \bibnamefont{West}},
  \bibinfo{journal}{Phys. Rev. Lett.} \textbf{\bibinfo{volume}{68}},
  \bibinfo{pages}{3623} (\bibinfo{year}{1992}).

\bibitem[{\citenamefont{Pinczuk et~al.}(1993)\citenamefont{Pinczuk, Dennis,
  Pfeiffer, and West}}]{Pinczuk1993}
\bibinfo{author}{\bibfnamefont{A.}~\bibnamefont{Pinczuk}},
  \bibinfo{author}{\bibfnamefont{B.~S.} \bibnamefont{Dennis}},
  \bibinfo{author}{\bibfnamefont{L.~N.} \bibnamefont{Pfeiffer}},
  \bibnamefont{and} \bibinfo{author}{\bibfnamefont{K.}~\bibnamefont{West}},
  \bibinfo{journal}{Phys. Rev. Lett.} \textbf{\bibinfo{volume}{70}},
  \bibinfo{pages}{3983} (\bibinfo{year}{1993}).

\bibitem[{\citenamefont{Davies et~al.}(1997)\citenamefont{Davies, Harris, Ryan,
  and Turberfield}}]{Davies1997}
\bibinfo{author}{\bibfnamefont{H.~D.~M.} \bibnamefont{Davies}},
  \bibinfo{author}{\bibfnamefont{J.~C.} \bibnamefont{Harris}},
  \bibinfo{author}{\bibfnamefont{J.~F.} \bibnamefont{Ryan}}, \bibnamefont{and}
  \bibinfo{author}{\bibfnamefont{A.~J.} \bibnamefont{Turberfield}},
  \bibinfo{journal}{Phys. Rev. Lett.} \textbf{\bibinfo{volume}{78}},
  \bibinfo{pages}{4095} (\bibinfo{year}{1997}).

\bibitem[{\citenamefont{Schuller et~al.}(1997)\citenamefont{Schuller, Krahne,
  Biese, Steinebach, Ulrichs, and Heitmann}}]{Schuller1997}
\bibinfo{author}{\bibfnamefont{C.}~\bibnamefont{Schuller}},
  \bibinfo{author}{\bibfnamefont{R.}~\bibnamefont{Krahne}},
  \bibinfo{author}{\bibfnamefont{G.}~\bibnamefont{Biese}},
  \bibinfo{author}{\bibfnamefont{C.}~\bibnamefont{Steinebach}},
  \bibinfo{author}{\bibfnamefont{E.}~\bibnamefont{Ulrichs}}, \bibnamefont{and}
  \bibinfo{author}{\bibfnamefont{D.}~\bibnamefont{Heitmann}},
  \bibinfo{journal}{Phys. Rev. B} \textbf{\bibinfo{volume}{56}},
  \bibinfo{pages}{1037} (\bibinfo{year}{1997}).

\bibitem[{\citenamefont{Pellegrini et~al.}(1998)\citenamefont{Pellegrini,
  Pinczuk, Dennis, Plaut, Pfeiffer, and West}}]{Pellegrini1998}
\bibinfo{author}{\bibfnamefont{V.}~\bibnamefont{Pellegrini}},
  \bibinfo{author}{\bibfnamefont{A.}~\bibnamefont{Pinczuk}},
  \bibinfo{author}{\bibfnamefont{B.~S.} \bibnamefont{Dennis}},
  \bibinfo{author}{\bibfnamefont{A.~S.} \bibnamefont{Plaut}},
  \bibinfo{author}{\bibfnamefont{L.~N.} \bibnamefont{Pfeiffer}},
  \bibnamefont{and} \bibinfo{author}{\bibfnamefont{K.~W.} \bibnamefont{West}},
  \bibinfo{journal}{Science} \textbf{\bibinfo{volume}{281}},
  \bibinfo{pages}{799} (\bibinfo{year}{1998}).

\bibitem[{\citenamefont{Eriksson et~al.}(1999)\citenamefont{Eriksson, Pinczuk,
  Dennis, Pfeiffer, and West}}]{Eriksson1999}
\bibinfo{author}{\bibfnamefont{M.~A.} \bibnamefont{Eriksson}},
  \bibinfo{author}{\bibfnamefont{A.}~\bibnamefont{Pinczuk}},
  \bibinfo{author}{\bibfnamefont{B.~S.} \bibnamefont{Dennis}},
  \bibinfo{author}{\bibfnamefont{L.~N.} \bibnamefont{Pfeiffer}},
  \bibnamefont{and} \bibinfo{author}{\bibfnamefont{K.~W.} \bibnamefont{West}},
  \bibinfo{journal}{Phys. Rev. Lett} \textbf{\bibinfo{volume}{82}},
  \bibinfo{pages}{2163} (\bibinfo{year}{1999}).

\bibitem[{\citenamefont{Kang et~al.}(2001)\citenamefont{Kang, Pinczuk, Dennis,
  Pfeiffer, and West}}]{Kang2001}
\bibinfo{author}{\bibfnamefont{M.}~\bibnamefont{Kang}},
  \bibinfo{author}{\bibfnamefont{A.}~\bibnamefont{Pinczuk}},
  \bibinfo{author}{\bibfnamefont{B.~S.} \bibnamefont{Dennis}},
  \bibinfo{author}{\bibfnamefont{L.~N.} \bibnamefont{Pfeiffer}},
  \bibnamefont{and} \bibinfo{author}{\bibfnamefont{K.~W.} \bibnamefont{West}},
  \bibinfo{journal}{Phys. Rev. Lett.} \textbf{\bibinfo{volume}{86}},
  \bibinfo{pages}{2637} (\bibinfo{year}{2001}).

\bibitem[{\citenamefont{Kulik et~al.}(2001)\citenamefont{Kulik, Kukushkin,
  Kirpichev, Smet, v.~Klitzing, and Wegscheider}}]{Kulik2001}
\bibinfo{author}{\bibfnamefont{L.~V.} \bibnamefont{Kulik}},
  \bibinfo{author}{\bibfnamefont{I.~V.} \bibnamefont{Kukushkin}},
  \bibinfo{author}{\bibfnamefont{V.~E.} \bibnamefont{Kirpichev}},
  \bibinfo{author}{\bibfnamefont{J.~H.} \bibnamefont{Smet}},
  \bibinfo{author}{\bibfnamefont{K.}~\bibnamefont{v.~Klitzing}},
  \bibnamefont{and}
  \bibinfo{author}{\bibfnamefont{W.}~\bibnamefont{Wegscheider}},
  \bibinfo{journal}{Phys. Rev. B} \textbf{\bibinfo{volume}{63}},
  \bibinfo{pages}{201402} (\bibinfo{year}{2001}).

\bibitem[{\citenamefont{Dujovne
  et~al.}(2003{\natexlab{a}})\citenamefont{Dujovne, Hirjibehedin, Pinczuk,
  Kang, Dennis, Pfeiffer, and West}}]{Dujovne2003b}
\bibinfo{author}{\bibfnamefont{I.}~\bibnamefont{Dujovne}},
  \bibinfo{author}{\bibfnamefont{C.}~\bibnamefont{Hirjibehedin}},
  \bibinfo{author}{\bibfnamefont{A.}~\bibnamefont{Pinczuk}},
  \bibinfo{author}{\bibfnamefont{M.}~\bibnamefont{Kang}},
  \bibinfo{author}{\bibfnamefont{B.}~\bibnamefont{Dennis}},
  \bibinfo{author}{\bibfnamefont{L.}~\bibnamefont{Pfeiffer}}, \bibnamefont{and}
  \bibinfo{author}{\bibfnamefont{K.}~\bibnamefont{West}},
  \bibinfo{journal}{Solid State Comm.}  (\bibinfo{year}{2003}{\natexlab{a}}),
  \bibinfo{note}{to be published, cond-mat/0306179}.

\bibitem[{\citenamefont{Burstein et~al.}(1979)\citenamefont{Burstein, Pinczuk,
  and Buchner}}]{Burstein1979}
\bibinfo{author}{\bibfnamefont{E.}~\bibnamefont{Burstein}},
  \bibinfo{author}{\bibfnamefont{A.}~\bibnamefont{Pinczuk}}, \bibnamefont{and}
  \bibinfo{author}{\bibfnamefont{S.}~\bibnamefont{Buchner}},
  \emph{\bibinfo{title}{Physics of Semiconductors 1978}}
  (\bibinfo{publisher}{Institute of Physics, London}, \bibinfo{year}{1979}).

\bibitem[{\citenamefont{Burstein et~al.}(1980)\citenamefont{Burstein, Pinczuk,
  and Millis}}]{Burstein1980}
\bibinfo{author}{\bibfnamefont{E.}~\bibnamefont{Burstein}},
  \bibinfo{author}{\bibfnamefont{A.}~\bibnamefont{Pinczuk}}, \bibnamefont{and}
  \bibinfo{author}{\bibfnamefont{D.~L.} \bibnamefont{Millis}},
  \bibinfo{journal}{Surf. Sci.} \textbf{\bibinfo{volume}{98}},
  \bibinfo{pages}{451} (\bibinfo{year}{1980}).

\bibitem[{\citenamefont{Pinczuk
  et~al.}(1994{\natexlab{a}})\citenamefont{Pinczuk, Dennis, Pfeiffer, West, and
  Burstein}}]{Pinczuk1994}
\bibinfo{author}{\bibfnamefont{A.}~\bibnamefont{Pinczuk}},
  \bibinfo{author}{\bibfnamefont{B.~S.} \bibnamefont{Dennis}},
  \bibinfo{author}{\bibfnamefont{L.~N.} \bibnamefont{Pfeiffer}},
  \bibinfo{author}{\bibfnamefont{K.~W.} \bibnamefont{West}}, \bibnamefont{and}
  \bibinfo{author}{\bibfnamefont{E.}~\bibnamefont{Burstein}},
  \bibinfo{journal}{Phil. Mag. B} \textbf{\bibinfo{volume}{70}},
  \bibinfo{pages}{429} (\bibinfo{year}{1994}{\natexlab{a}}).

\bibitem[{\citenamefont{Platzman and He}(1994)}]{Platzman1994}
\bibinfo{author}{\bibfnamefont{P.~M.} \bibnamefont{Platzman}} \bibnamefont{and}
  \bibinfo{author}{\bibfnamefont{S.}~\bibnamefont{He}}, \bibinfo{journal}{Phys.
  Rev. B} \textbf{\bibinfo{volume}{49}}, \bibinfo{pages}{13674}
  (\bibinfo{year}{1994}).

\bibitem[{\citenamefont{{Das Sarma} and Wang}(1999)}]{DasSarma1999}
\bibinfo{author}{\bibfnamefont{S.}~\bibnamefont{{Das Sarma}}} \bibnamefont{and}
  \bibinfo{author}{\bibfnamefont{D.-W.} \bibnamefont{Wang}},
  \bibinfo{journal}{Phys. Rev. Lett.} \textbf{\bibinfo{volume}{83}},
  \bibinfo{pages}{816} (\bibinfo{year}{1999}).

\bibitem[{\citenamefont{Fromer et~al.}(2002)\citenamefont{Fromer, Lai, Chemla,
  Perakis, Driscoll, and Gossard}}]{Fromer2002}
\bibinfo{author}{\bibfnamefont{N.~A.} \bibnamefont{Fromer}},
  \bibinfo{author}{\bibfnamefont{C.~E.} \bibnamefont{Lai}},
  \bibinfo{author}{\bibfnamefont{D.~S.} \bibnamefont{Chemla}},
  \bibinfo{author}{\bibfnamefont{I.~E.} \bibnamefont{Perakis}},
  \bibinfo{author}{\bibfnamefont{D.}~\bibnamefont{Driscoll}}, \bibnamefont{and}
  \bibinfo{author}{\bibfnamefont{A.~C.} \bibnamefont{Gossard}},
  \bibinfo{journal}{Phys. Rev. Lett.} \textbf{\bibinfo{volume}{89}},
  \bibinfo{pages}{067401} (\bibinfo{year}{2002}).

\bibitem[{\citenamefont{Bao et~al.}(2002)\citenamefont{Bao, Pfeiffer, West, and
  Merlin}}]{Bao2002}
\bibinfo{author}{\bibfnamefont{J.}~\bibnamefont{Bao}},
  \bibinfo{author}{\bibfnamefont{L.~N.} \bibnamefont{Pfeiffer}},
  \bibinfo{author}{\bibfnamefont{K.~W.} \bibnamefont{West}}, \bibnamefont{and}
  \bibinfo{author}{\bibfnamefont{R.}~\bibnamefont{Merlin}},
  \bibinfo{journal}{QELS 2002 Technical Digest} \textbf{\bibinfo{volume}{74}},
  \bibinfo{pages}{261} (\bibinfo{year}{2002}).

\bibitem[{\citenamefont{Marmorkos and {Das
  Sarma}}(1992)}]{Marmorkos&DasSarma1992}
\bibinfo{author}{\bibfnamefont{I.~K.} \bibnamefont{Marmorkos}}
  \bibnamefont{and} \bibinfo{author}{\bibfnamefont{S.}~\bibnamefont{{Das
  Sarma}}}, \bibinfo{journal}{Phys. Rev. B} \textbf{\bibinfo{volume}{45}},
  \bibinfo{pages}{13396} (\bibinfo{year}{1992}).

\bibitem[{\citenamefont{MacDonald et~al.}(1992)\citenamefont{MacDonald, Rezayi,
  and Keller}}]{MacDonald1992}
\bibinfo{author}{\bibfnamefont{A.~H.} \bibnamefont{MacDonald}},
  \bibinfo{author}{\bibfnamefont{E.~H.} \bibnamefont{Rezayi}},
  \bibnamefont{and} \bibinfo{author}{\bibfnamefont{D.}~\bibnamefont{Keller}},
  \bibinfo{journal}{Phys. Rev. Lett.} \textbf{\bibinfo{volume}{68}},
  \bibinfo{pages}{1939} (\bibinfo{year}{1992}).

\bibitem[{\citenamefont{Wang et~al.}(1992)\citenamefont{Wang, Birman, and
  Su}}]{Wang1992}
\bibinfo{author}{\bibfnamefont{B.-S.} \bibnamefont{Wang}},
  \bibinfo{author}{\bibfnamefont{J.~L.} \bibnamefont{Birman}},
  \bibnamefont{and} \bibinfo{author}{\bibfnamefont{Z.-B.} \bibnamefont{Su}},
  \bibinfo{journal}{Phys. Rev. Lett.} \textbf{\bibinfo{volume}{68}},
  \bibinfo{pages}{1605} (\bibinfo{year}{1992}).

\bibitem[{\citenamefont{Apalkov and Rashba}(1993)}]{Apalkov1993}
\bibinfo{author}{\bibfnamefont{V.~M.} \bibnamefont{Apalkov}} \bibnamefont{and}
  \bibinfo{author}{\bibfnamefont{E.~I.} \bibnamefont{Rashba}},
  \bibinfo{journal}{Phys. Rev. B} \textbf{\bibinfo{volume}{48}},
  \bibinfo{pages}{18312} (\bibinfo{year}{1993}).

\bibitem[{\citenamefont{Kheng et~al.}(1993)\citenamefont{Kheng, Cox, {d\'
  Aubigne}, Bassani, Saminadayar, and Tatarenko}}]{Kheng1993}
\bibinfo{author}{\bibfnamefont{K.}~\bibnamefont{Kheng}},
  \bibinfo{author}{\bibfnamefont{R.~T.} \bibnamefont{Cox}},
  \bibinfo{author}{\bibfnamefont{M.~Y.} \bibnamefont{{d\' Aubigne}}},
  \bibinfo{author}{\bibfnamefont{F.}~\bibnamefont{Bassani}},
  \bibinfo{author}{\bibfnamefont{K.}~\bibnamefont{Saminadayar}},
  \bibnamefont{and}
  \bibinfo{author}{\bibfnamefont{S.}~\bibnamefont{Tatarenko}},
  \bibinfo{journal}{Phys. Rev. Lett.} \textbf{\bibinfo{volume}{71}},
  \bibinfo{pages}{1752} (\bibinfo{year}{1993}).

\bibitem[{\citenamefont{Wojs and Hawrylak}(1995)}]{Wojs1995}
\bibinfo{author}{\bibfnamefont{A.}~\bibnamefont{Wojs}} \bibnamefont{and}
  \bibinfo{author}{\bibfnamefont{P.}~\bibnamefont{Hawrylak}},
  \bibinfo{journal}{Phys. Rev. B} \textbf{\bibinfo{volume}{51}},
  \bibinfo{pages}{10880} (\bibinfo{year}{1995}).

\bibitem[{\citenamefont{Finkelstein et~al.}(1995)\citenamefont{Finkelstein,
  Shtrikman, and Bar-Joseph}}]{Finkelstein1995}
\bibinfo{author}{\bibfnamefont{G.}~\bibnamefont{Finkelstein}},
  \bibinfo{author}{\bibfnamefont{H.}~\bibnamefont{Shtrikman}},
  \bibnamefont{and}
  \bibinfo{author}{\bibfnamefont{I.}~\bibnamefont{Bar-Joseph}},
  \bibinfo{journal}{Phys. Rev. Lett.} \textbf{\bibinfo{volume}{74}},
  \bibinfo{pages}{976} (\bibinfo{year}{1995}).

\bibitem[{\citenamefont{Shields et~al.}(1995)\citenamefont{Shields, Pepper,
  Simmons, and Ritchie}}]{Shields1995}
\bibinfo{author}{\bibfnamefont{A.~J.} \bibnamefont{Shields}},
  \bibinfo{author}{\bibfnamefont{M.}~\bibnamefont{Pepper}},
  \bibinfo{author}{\bibfnamefont{M.~Y.} \bibnamefont{Simmons}},
  \bibnamefont{and} \bibinfo{author}{\bibfnamefont{D.~A.}
  \bibnamefont{Ritchie}}, \bibinfo{journal}{Phys. Rev. B}
  \textbf{\bibinfo{volume}{52}}, \bibinfo{pages}{7841} (\bibinfo{year}{1995}).

\bibitem[{\citenamefont{Palacios et~al.}(1996)\citenamefont{Palacios, Yoshioka,
  and MacDonald}}]{Palacios1996}
\bibinfo{author}{\bibfnamefont{J.~J.} \bibnamefont{Palacios}},
  \bibinfo{author}{\bibfnamefont{D.}~\bibnamefont{Yoshioka}}, \bibnamefont{and}
  \bibinfo{author}{\bibfnamefont{A.~H.} \bibnamefont{MacDonald}},
  \bibinfo{journal}{Phys. Rev. B} \textbf{\bibinfo{volume}{54}},
  \bibinfo{pages}{R2296} (\bibinfo{year}{1996}).

\bibitem[{\citenamefont{Gekhtman et~al.}(1997)\citenamefont{Gekhtman, Cohen,
  Ron, and Pfeiffer}}]{Gekhtman1997}
\bibinfo{author}{\bibfnamefont{D.}~\bibnamefont{Gekhtman}},
  \bibinfo{author}{\bibfnamefont{E.}~\bibnamefont{Cohen}},
  \bibinfo{author}{\bibfnamefont{A.}~\bibnamefont{Ron}}, \bibnamefont{and}
  \bibinfo{author}{\bibfnamefont{L.~N.} \bibnamefont{Pfeiffer}},
  \bibinfo{journal}{Phys. Rev. B} \textbf{\bibinfo{volume}{56}},
  \bibinfo{pages}{12768} (\bibinfo{year}{1997}).

\bibitem[{\citenamefont{Wojs et~al.}(1999)\citenamefont{Wojs, Szlufarska, Yi,
  and Quinn}}]{Wojs1999}
\bibinfo{author}{\bibfnamefont{A.}~\bibnamefont{Wojs}},
  \bibinfo{author}{\bibfnamefont{I.}~\bibnamefont{Szlufarska}},
  \bibinfo{author}{\bibfnamefont{K.~S.} \bibnamefont{Yi}}, \bibnamefont{and}
  \bibinfo{author}{\bibfnamefont{J.~J.} \bibnamefont{Quinn}},
  \bibinfo{journal}{Phys. Rev. B} \textbf{\bibinfo{volume}{60}},
  \bibinfo{pages}{11273} (\bibinfo{year}{1999}).

\bibitem[{\citenamefont{Wojs et~al.}(2000)\citenamefont{Wojs, Quinn, and
  Hawrylak}}]{Wojs2000}
\bibinfo{author}{\bibfnamefont{A.}~\bibnamefont{Wojs}},
  \bibinfo{author}{\bibfnamefont{J.~J.} \bibnamefont{Quinn}}, \bibnamefont{and}
  \bibinfo{author}{\bibfnamefont{P.}~\bibnamefont{Hawrylak}},
  \bibinfo{journal}{Phys. Rev. B} \textbf{\bibinfo{volume}{62}},
  \bibinfo{pages}{4630} (\bibinfo{year}{2000}).

\bibitem[{\citenamefont{Yusa et~al.}(2001)\citenamefont{Yusa, Shtrikman, and
  Bar-Joseph}}]{Yusa2001}
\bibinfo{author}{\bibfnamefont{G.}~\bibnamefont{Yusa}},
  \bibinfo{author}{\bibfnamefont{H.}~\bibnamefont{Shtrikman}},
  \bibnamefont{and}
  \bibinfo{author}{\bibfnamefont{I.}~\bibnamefont{Bar-Joseph}},
  \bibinfo{journal}{Phys. Rev. Lett.} \textbf{\bibinfo{volume}{87}},
  \bibinfo{pages}{216402} (\bibinfo{year}{2001}).

\bibitem[{\citenamefont{Schuller et~al.}(2001)\citenamefont{Schuller, Broocks,
  Heyn, and Heitmann}}]{Schuller2001}
\bibinfo{author}{\bibfnamefont{C.}~\bibnamefont{Schuller}},
  \bibinfo{author}{\bibfnamefont{K.}~\bibnamefont{Broocks}},
  \bibinfo{author}{\bibfnamefont{C.}~\bibnamefont{Heyn}}, \bibnamefont{and}
  \bibinfo{author}{\bibfnamefont{D.}~\bibnamefont{Heitmann}},
  \bibinfo{journal}{Phys. Rev. B} \textbf{\bibinfo{volume}{65}},
  \bibinfo{pages}{081301} (\bibinfo{year}{2001}).

\bibitem[{\citenamefont{Szlufarska et~al.}(2001)\citenamefont{Szlufarska, Wojs,
  and Quinn}}]{Szlufarska2001}
\bibinfo{author}{\bibfnamefont{I.}~\bibnamefont{Szlufarska}},
  \bibinfo{author}{\bibfnamefont{A.}~\bibnamefont{Wojs}}, \bibnamefont{and}
  \bibinfo{author}{\bibfnamefont{J.~J.} \bibnamefont{Quinn}},
  \bibinfo{journal}{Phys. Rev. B} \textbf{\bibinfo{volume}{63}},
  \bibinfo{pages}{085305} (\bibinfo{year}{2001}).

\bibitem[{\citenamefont{Nickel et~al.}(2002)\citenamefont{Nickel, Yeo,
  Dzyubenko, McCombe, Petrou, Sivachenko, Schaff, and Umansky}}]{Nickel2002}
\bibinfo{author}{\bibfnamefont{H.~A.} \bibnamefont{Nickel}},
  \bibinfo{author}{\bibfnamefont{T.~M.} \bibnamefont{Yeo}},
  \bibinfo{author}{\bibfnamefont{A.~B.} \bibnamefont{Dzyubenko}},
  \bibinfo{author}{\bibfnamefont{B.~D.} \bibnamefont{McCombe}},
  \bibinfo{author}{\bibfnamefont{A.}~\bibnamefont{Petrou}},
  \bibinfo{author}{\bibfnamefont{A.~Y.} \bibnamefont{Sivachenko}},
  \bibinfo{author}{\bibfnamefont{W.}~\bibnamefont{Schaff}}, \bibnamefont{and}
  \bibinfo{author}{\bibfnamefont{V.}~\bibnamefont{Umansky}},
  \bibinfo{journal}{Phys. Rev. Lett.} \textbf{\bibinfo{volume}{88}},
  \bibinfo{pages}{056801} (\bibinfo{year}{2002}).

\bibitem[{\citenamefont{Broocks et~al.}(2002)\citenamefont{Broocks, Schroter,
  Heitmann, Heyn, Schuller, Bichler, and Wegscheider}}]{Broocks2002}
\bibinfo{author}{\bibfnamefont{K.-B.} \bibnamefont{Broocks}},
  \bibinfo{author}{\bibfnamefont{P.}~\bibnamefont{Schroter}},
  \bibinfo{author}{\bibfnamefont{D.}~\bibnamefont{Heitmann}},
  \bibinfo{author}{\bibfnamefont{C.}~\bibnamefont{Heyn}},
  \bibinfo{author}{\bibfnamefont{C.}~\bibnamefont{Schuller}},
  \bibinfo{author}{\bibfnamefont{M.}~\bibnamefont{Bichler}}, \bibnamefont{and}
  \bibinfo{author}{\bibfnamefont{W.}~\bibnamefont{Wegscheider}},
  \bibinfo{journal}{Phys. Rev. B} \textbf{\bibinfo{volume}{66}},
  \bibinfo{pages}{041309} (\bibinfo{year}{2002}).

\bibitem[{\citenamefont{Kang et~al.}(2000)\citenamefont{Kang, Pinczuk, Dennis,
  Eriksson, Pfeiffer, and West}}]{Kang2000}
\bibinfo{author}{\bibfnamefont{M.}~\bibnamefont{Kang}},
  \bibinfo{author}{\bibfnamefont{A.}~\bibnamefont{Pinczuk}},
  \bibinfo{author}{\bibfnamefont{B.~S.} \bibnamefont{Dennis}},
  \bibinfo{author}{\bibfnamefont{M.~A.} \bibnamefont{Eriksson}},
  \bibinfo{author}{\bibfnamefont{L.~N.} \bibnamefont{Pfeiffer}},
  \bibnamefont{and} \bibinfo{author}{\bibfnamefont{K.~W.} \bibnamefont{West}},
  \bibinfo{journal}{Phys. Rev. Lett.} \textbf{\bibinfo{volume}{84}},
  \bibinfo{pages}{546} (\bibinfo{year}{2000}).

\bibitem[{\citenamefont{Dujovne
  et~al.}(2003{\natexlab{b}})\citenamefont{Dujovne, Pinczuk, Kang, Dennis,
  Pfeiffer, and West}}]{Dujovne2003a}
\bibinfo{author}{\bibfnamefont{I.}~\bibnamefont{Dujovne}},
  \bibinfo{author}{\bibfnamefont{A.}~\bibnamefont{Pinczuk}},
  \bibinfo{author}{\bibfnamefont{M.}~\bibnamefont{Kang}},
  \bibinfo{author}{\bibfnamefont{B.~S.} \bibnamefont{Dennis}},
  \bibinfo{author}{\bibfnamefont{L.~N.} \bibnamefont{Pfeiffer}},
  \bibnamefont{and} \bibinfo{author}{\bibfnamefont{K.~W.} \bibnamefont{West}},
  \bibinfo{journal}{Phys. Rev. Lett} \textbf{\bibinfo{volume}{90}},
  \bibinfo{pages}{036803} (\bibinfo{year}{2003}{\natexlab{b}}).

\bibitem[{\citenamefont{Hirjibehedin et~al.}(2003)\citenamefont{Hirjibehedin,
  Pinczuk, Dennis, Pfeiffer, and West}}]{Hirjibehedin2003}
\bibinfo{author}{\bibfnamefont{C.~F.} \bibnamefont{Hirjibehedin}},
  \bibinfo{author}{\bibfnamefont{A.}~\bibnamefont{Pinczuk}},
  \bibinfo{author}{\bibfnamefont{B.~S.} \bibnamefont{Dennis}},
  \bibinfo{author}{\bibfnamefont{L.~N.} \bibnamefont{Pfeiffer}},
  \bibnamefont{and} \bibinfo{author}{\bibfnamefont{K.~W.} \bibnamefont{West}},
  \bibinfo{journal}{cond-mat/0306152}  (\bibinfo{year}{2003}).

\bibitem[{\citenamefont{Pinczuk
  et~al.}(1994{\natexlab{b}})\citenamefont{Pinczuk, Dennis, Pfeiffer, and
  West}}]{PinczukSST1994}
\bibinfo{author}{\bibfnamefont{A.}~\bibnamefont{Pinczuk}},
  \bibinfo{author}{\bibfnamefont{B.~S.} \bibnamefont{Dennis}},
  \bibinfo{author}{\bibfnamefont{L.~N.} \bibnamefont{Pfeiffer}},
  \bibnamefont{and} \bibinfo{author}{\bibfnamefont{K.~W.} \bibnamefont{West}},
  \bibinfo{journal}{Semicond. Sci. Technol.} \textbf{\bibinfo{volume}{9}},
  \bibinfo{pages}{1865} (\bibinfo{year}{1994}{\natexlab{b}}).

\bibitem[{\citenamefont{Nakajima and Aoki}(1994)}]{Nakajima&Aoki1994}
\bibinfo{author}{\bibfnamefont{T.}~\bibnamefont{Nakajima}} \bibnamefont{and}
  \bibinfo{author}{\bibfnamefont{H.}~\bibnamefont{Aoki}},
  \bibinfo{journal}{Phys. Rev. Lett.} \textbf{\bibinfo{volume}{73}},
  \bibinfo{pages}{3568} (\bibinfo{year}{1994}).

\bibitem[{\citenamefont{Pinczuk et~al.}(1998)\citenamefont{Pinczuk, Dennis,
  Pfeiffer, and West}}]{Pinczuk1998}
\bibinfo{author}{\bibfnamefont{A.}~\bibnamefont{Pinczuk}},
  \bibinfo{author}{\bibfnamefont{B.~S.} \bibnamefont{Dennis}},
  \bibinfo{author}{\bibfnamefont{L.~N.} \bibnamefont{Pfeiffer}},
  \bibnamefont{and} \bibinfo{author}{\bibfnamefont{K.~W.} \bibnamefont{West}},
  \bibinfo{journal}{Physica B} \textbf{\bibinfo{volume}{249}},
  \bibinfo{pages}{40} (\bibinfo{year}{1998}).

\bibitem[{\citenamefont{Danan et~al.}(1989)\citenamefont{Danan, Pinczuk,
  Valladares, Pfeiffer, West, and Tu}}]{Danan1989}
\bibinfo{author}{\bibfnamefont{G.}~\bibnamefont{Danan}},
  \bibinfo{author}{\bibfnamefont{A.}~\bibnamefont{Pinczuk}},
  \bibinfo{author}{\bibfnamefont{J.~P.} \bibnamefont{Valladares}},
  \bibinfo{author}{\bibfnamefont{L.~N.} \bibnamefont{Pfeiffer}},
  \bibinfo{author}{\bibfnamefont{K.~W.} \bibnamefont{West}}, \bibnamefont{and}
  \bibinfo{author}{\bibfnamefont{C.~W.} \bibnamefont{Tu}},
  \bibinfo{journal}{Phys. Rev. B} \textbf{\bibinfo{volume}{39}},
  \bibinfo{pages}{5512} (\bibinfo{year}{1989}).

\bibitem[{\citenamefont{Mandal and Jain}(2001)}]{Mandal2001}
\bibinfo{author}{\bibfnamefont{S.~S.} \bibnamefont{Mandal}} \bibnamefont{and}
  \bibinfo{author}{\bibfnamefont{J.~K.} \bibnamefont{Jain}},
  \bibinfo{journal}{Phys. Rev. B} \textbf{\bibinfo{volume}{64}},
  \bibinfo{pages}{125310} (\bibinfo{year}{2001}).

\end{thebibliography}

\end{document}